\documentclass[aps,prb,amsmath,amssymb,twocolumn,superscriptaddress,showpacs,longbibliography,floatfix]{revtex4-1}
\usepackage{amsmath,amssymb}
\usepackage{graphicx}
\usepackage{appendix}
\usepackage[usenames,dvipsnames]{color}
\usepackage[colorlinks=true, citecolor=blue, linkcolor=blue, urlcolor=blue]{hyperref}
\usepackage{bm}
\usepackage[nottoc,numbib]{tocbibind}
\usepackage{algpseudocode} 
\usepackage{ulem}

\newcommand{\vect}[1]{\boldsymbol{#1}}
\newcommand{\kk}{\vect{k}}

\newcommand*{\affilmpsd}{Max Planck Institute for the Structure and Dynamics of Matter, Center for Free Electron Laser Science (CFEL), Luruper Chaussee 149, 22761 Hamburg, Germany}

\begin{document}

\title{Nonequilibrium phase transition in a driven-dissipative quantum antiferromagnet}

\author{Mona H.~Kalthoff}
\email{mona.kalthoff@mpsd.mpg.de}
\affiliation{\affilmpsd}

\author{Dante M.~Kennes}
\email{dante.kennes@rwth-aachen.de}
\affiliation 
{Institut f\"ur Theorie der Statistischen Physik, RWTH Aachen University, 52056 Aachen, Germany and JARA-Fundamentals of Future Information Technology, 52056 Aachen, Germany}
\affiliation{\affilmpsd}

\author{Andrew J.~Millis}
\email{amillis@flatironinstitute.org}
\affiliation 
{Department of Physics, Columbia University, 538 West 120th Street, New York, New York 10027, USA}
\affiliation 
{Center for Computational Quantum Physics, Flatiron Institute, 162 5th Avenue, New York, New York 10010, USA}

\author{Michael A.~Sentef}
\email{michael.sentef@mpsd.mpg.de}
\affiliation{\affilmpsd}

\date{\today}

\begin{abstract}
   A deeper theoretical understanding of driven-dissipative interacting systems and their nonequilibrium phase transitions is essential both to advance our fundamental physics understanding and to harness technological opportunities arising from optically controlled quantum many-body states. This paper provides a numerical study of dynamical phases and the transitions between them 
   in the nonequilibrium steady state of  the prototypical two-dimensional Heisenberg antiferromagnet with drive and dissipation. 
   We demonstrate a nonthermal transition that is characterized by a qualitative change in the magnon distribution, from subthermal at low drive to a generalized Bose-Einstein form including a nonvanishing condensate fraction at high drive. A finite-size analysis reveals static and dynamical critical scaling at  the transition, with a discontinuous slope of the magnon number versus driving field strength and critical slowing down at the transition point. Implications for experiments on quantum materials and polariton condensates are discussed.
\end{abstract}
\maketitle

\section{Introduction}
Nonequilibrium phase transitions in driven interacting quantum systems constitute a fundamental and largely open research problem \cite{basov_towards_2017,de_la_torre_nonthermal_2021}. \textit{Quenches}, i.e., abrupt changes in Hamiltonian parameters 
  or initial conditions, followed by a time evolution, have been extensively studied and can lead to dynamical phase transitions~\cite{Tsuji_2013, Klinder_2015_Dynamical_phase_transition_in_the_open_Dicke_model} characterized by qualitative modifications of the dynamical response as the quench magnitude is varied. A nonequilibrium steady state presents additional issues involving the flow and redistribution of energy: the drive adds energy, the dissipation removes energy, and the internal dynamics redistribute energy among modes ~\cite{Kemper_general_principles_2018,Yarmohammadi_2021_Dynamical_properties_of_driven_dissipative}. As the drive strength is varied, the competition between these effects can qualitatively change system properties in the same sense that changing temperature or a Hamiltonian parameter can drive a system through an equilibrium phase transition.  

Equilibrium phase transitions are typically analyzed in terms of the onset or disappearance of order parameters that encode broken symmetries, for example, the staggered magnetization in an antiferromagnet that appears when the temperature is reduced below a critical temperature. We label such  phase transitions as symmetry breaking transition 
in the following. In a nonequilibrium setting, an  additional type of phase transition can exist that is characterized by a qualitative change in the low-frequency distribution of the collective excitations of a system. Such a transition cannot exist in equilibrium where the form of the distribution is fixed by equilibrium thermodynamics.  We refer to the latter as a subthermal-to-superthermal transition. 
Phase transitions occurring in a nonequilibrium \textit{steady state} are the subject of an interesting and growing literature~\cite{Maghrebi_Nonequilibrium_many-body_2016, Millis_2006_Nonequilibrium_Quantum_Criticality, Millis_2007_Coulomb_Keldysh_contour, Millis_2008_quantum_criticality_ferromagnets, Millis_2011_Current_driven_transition, brennecke_real-time_2013,Rota_2019,BKTnoneq, Marino_Driven_Markovian_Quantum_Criticality_2016} but are less well understood. A deeper theoretical understanding of these issues could open nonthermal pathways for controlling emergent properties of driven quantum materials \cite{de_la_torre_nonthermal_2021}.

Driven magnetic systems are of particular interest in this context, for both fundamental and technological reasons \cite{Barman_Magnonic_Roadmap_2021}. A specific focus of  attention has been the possibility of magnon Bose-Einstein condensation (BEC), in which a system is excited by a radiation pulse and the resulting  excitation distribution forms a single coherent macroscopic quantum state with the lowest energy excited state being macroscopically populated. The existing  experimental literature on magnon BEC \cite{Demokritov_BE_condensation_2006, Bozhko_Supercurrent_2016, Nowik-Boltyk_Spatially_nonuniform_2012, Bender_dc_pumped_condensates_2014, Clausen_magnon_gas_2015, Demidov_Magnon_Kinetics_2008, Serga_YIG_magnonics_2010, Kreil_Tunable_space_time_2019, Sun_YIG_films_2017,Bunkov2011, Bunkov_Magnon_Bose_Einstein_condensation_spin_superfluidity_2010, Autti_Bose-Einstein_Condensation_of_Magnons_and_Spin_Superfluidity_2018, Bunkov_Magnon_Condensation_2007,Serga_Bose--Einstein_condensation_ultra-hot_2014, Kreil_Kinetic_Instability_2018, Melkov_Kinetic_instability_1994} concerns systems with very long energy relaxation times, where a population of magnons is transiently induced (often by a short duration frequency-coherent excitation) and then evolves into a BEC \cite{Zapf_Bose-Einstein_condensation_in_quantum_magnets_2014, Barman_Magnonic_Roadmap_2021, Bunkov_Magnon_condensation_2018, Pirro_Advances_coherent_magnonics_2021}.  This physics is very  similar to the Bose-Einstein condensation  of excitons and  exciton-polaritons which has been  studied experimentally \cite{Byrnes_2014_Exciton_polariton_condensates, Plumhof2014, Walker2018_Driven-dissipative_non-equilibrium_Bose-Einstein_condensation, hakala_boseeinstein_2018, Vakevainen2020} and theoretically \cite{Deng_Exciton-polariton_2010, sieberer_dynamical_2013, Sieberer_2014_Bose_condensation_polaritons}. Theoretical analyses of the magnon case to date have been based  on semi-phenomenological continuum approximations using Landau-Lifshitz-Gilbert equations \cite{Rueckriegel_Rayleigh-Jeans_Condensation_2015, Mohseni_magnons_confined_systems_2020},  Gross-Pitaevskii equations \cite{Bozhko_Supercurrent_2016, Bunkov_Magnon_Bose_Einstein_condensation_spin_superfluidity_2010, Bunkov2011} or field theoretical analyses  \cite{Millis_2006_Nonequilibrium_Quantum_Criticality,Millis_2008_quantum_criticality_ferromagnets,sieberer_dynamical_2013, Sieberer_2014_Bose_condensation_polaritons,brennecke_real-time_2013,Rota_2019,BKTnoneq, Marino_Driven_Markovian_Quantum_Criticality_2016}. Here, we focus on the distribution function of excitations.

\begin{figure}
	\centering
	\includegraphics[width= 0.9\columnwidth]{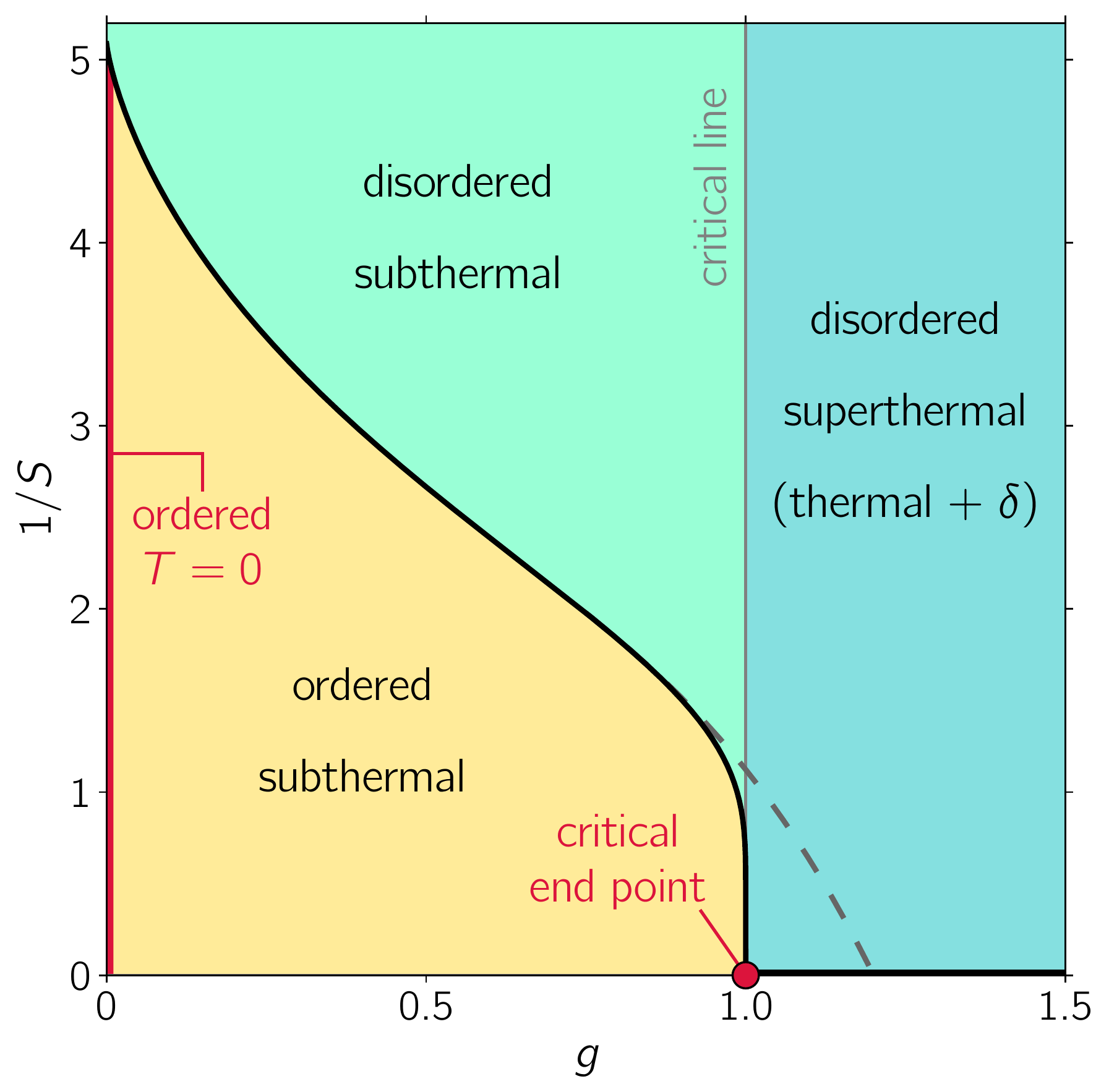}
	\caption{
	\textbf{Nonequilibrium phase diagram of driven dissipative steady states.} Steady states as a function of drive strength $g$ and quantum fluctuations, parametrized here by the inverse spin length $1/S$, but controlled in physical systems by many factors, including geometrical frustration. The red section along the vertical axis marks the antiferromagnetically ordered ground state at $T=0$. The black curve separating the ordered (orange) from the disordered (green) subthermal phase is obtained by determining the value of $1/S$ at which the staggered magnetization (as defined in Eq.~\eqref{eq:magnetization}) vanishes for a given $g$. The grey vertical line at $g=1$ separates the subthermal from the superthermal regime, which turns into a thermal distribution plus a $\delta$ function in the interacting system in the thermodynamic limit. The critical end point at $g=1$, $1/S=0$ is a specific feature of the Heisenberg antiferromagnet in two dimensions. The grey dashed curve indicates the expected behavior in three dimensions, or in the anisotropic xy or xxz (Ising, gapped) regimes in two dimensions.}
	\label{fig:1_Phase_diagram}
\end{figure}

In this work we aim to add a new dimension to the understanding of this field. We study a steady state system in which the crucial physics is the interplay of interactions and the flow of energy and particles from the drive through the system to a dissipative reservoir.  We provide a precise  microscopic treatment of the interaction among excitations, which is known  \cite{Zakharov_Spin-wave_turbulence_1975,Rezende_Coherence_microwave_driven_2009,Cornelissen_Magnon_Spin_transport_2016,Schneider_Rapid_Cooling_2020, Pirro_Advances_coherent_magnonics_2021,Alessio_2014_Long_time_Behavior_of_isolated_periodically_driven_systems, Tindall_2019_Heating-Induced_Long-Range_Hubbard_Model, Abanin_2017_prethermalization, Abanin2017_Many_Body_Prethermalization, Ho_2018_Prethermalization, Kuwahara_2016_transient_dynamics, Mori_2016_Energy_Absorbtion, Mori_2018_Thermalization_and_prethermalization} to be crucial for the long time physics. 
Fig.~\ref{fig:1_Phase_diagram} shows the behavior of the spin system under consideration as a function of the critical parameter $g$, which parametrizes the nonequilibrium excitation strength relative to dissipative losses and will be introduced in more detail below. The figure displays two distinct phase transitions, namely an order-to-disorder phase transition, which is conceptually similar to known equilibrium transitions but occurs here for nonthermal distributions, and an intrinsically nonequilibrium subthermal to superthermal transition, which we study in this paper. This new phase transition is characterized by a qualitative change in the distribution function.

\section{Model and Formalism}

\subsection{Hamiltonian and kinetic equation}
We study the driven-dissipative square-lattice Heisenberg antiferromagnet with nearest neighbor interactions, described by the  Hamiltonian
\begin{align}
\label{eq:Heisenberg_Hamiltonian}
H_{Heis}=J\sum_{\left\langle ij\right\rangle}\left\lbrace \frac{1}{2}
\left(S_i^+ S_j^- + S_i^- S_j^+\right)+ S_i^z S_j^z\right\rbrace,
\end{align}
with canonical spin operators $\bm{S}_i$ at site $i$ of the lattice.
The Heisenberg Hamiltonian has two parameters, the exchange coupling strength $J$, which sets the energy scale and which we take to be positive so that the ground state is antiferromagnetic, and the spin magnitude $|S|$ which sets the strength of the quantum fluctuations and of the interactions between the spin waves. At $|S|=\infty$ the model is straightforwardly solvable and has a two-fold degenerate set of spin wave excitations (magnons) with dispersion $\omega_{\vect{k}}$. The primary object of interest will be the magnon distribution function $n_{\vect{k}}$ counting the number of magnons excited above the ground state into the mode with energy $\omega_{\vect{k}}$. Key to our analysis will be the interactions between magnons.  Because we are interested  in the  qualitative effects of  the interactions we use a standard Holstein-Primakoff method~\cite{Holstein_Field_1940} to obtain the  spin-wave interactions at leading nontrivial order in $1/|S|$ (See appendix~\ref{sec:methods}).  The important points here are that the inter-spin-wave interactions conserve both total energy and the total number of spin waves and that their effect on the distribution may be studied using the Boltzmann equation with a collision integral $\mathcal{S}$ derived via standard methods from the magnon-magnon interactions. 

The Heisenberg model is an effective model describing the low energy physics of  a more fundamental system of strongly correlated electrons moving in a periodic lattice potential such as the Hubbard model. These more fundamental models enable a calculation of the  drive due to electromagnetic radiation and dissipation due to coupling with a reservoir.  We specifically adopt the model studied in Ref.~\onlinecite{walldorf_antiferromagnetic_2019} in which the Heisenberg model is obtained as the low-energy limit of the half-filled large $U$ Hubbard model. The drive emerges from a Floquet analysis of minimally coupled high frequency radiation detuned from the upper Hubbard band. The dissipation results from particle exchange with a reservoir, which we take to be at zero temperature. The particle exchange is virtual because of the Mott-Hubbard gap, but dissipation of energy and magnons into the reservoir are allowed.

Since we consider only a spatially uniform drive, we restrict our attention to a distribution function of energy $\omega$ (instead of momentum $\vect{k}$) defined \footnote{Here all integrals are understood as properly normalized over the magnetic Brillouin zone.} as 
$n(\omega)=\int d^2k n_{\vect{k}}\delta(\omega-\omega_{\vect{k}})/\rho(\omega)$ 
with $\omega_{\vect{k}}$ the magnon energy and $n_{\vect{k}} $ the magnon distribution as a function of wavevector. The density of states summed over the two magnon branches is 
\begin{equation}
\rho(\omega)=2\int d^2k \delta(\omega-\omega_{\vect{k}})\,.
\label{eq:DOS_continous}
\end{equation}
We take the drive and dissipation from a previous analysis \cite{walldorf_antiferromagnetic_2019} of the driven-dissipative Hubbard model, specializing to the particular case of a high-frequency drive detuned from any  charge excitations, and a dissipation arising from particle exchange with a reservoir. Reference~\cite{walldorf_antiferromagnetic_2019} found, using an approximation that neglected the magnon-magnon interactions,  that the effect of a high frequency detuned  drive is the addition of magnons to the system, such that the number of magnons in the mode with energy $\omega$ increases at the rate  $\gamma_{\mathrm{in}}(1+n(\omega))$. $\gamma_{in}$ is proportional to the drive strength and the simple form of the in-scattering follows from the very high frequency, detuned drive. The calculation also implies a decay of magnons into the charge reservoir at a rate given by $\gamma_{\mathrm{out}}\left(n(\omega)+\left(\frac{n(\omega)}{n_{\tilde{T}}(\omega)}\right)^2\right)$ with $n_{\tilde{T}}\left(\omega\right)=1/(e^\frac{\omega}{\tilde{T}}-1)$ and parameters $\tilde{T}\approx 0.6J$. Note that $\tilde{T}$ from Eq.~\eqref{eq:interacting_differential_equation} is not the equilibrium temperature of the system, but is a parameter describing the nonlinearity of the relaxation to the bath. 
The nonlinearity ensures a steady state at any drive amplitude. The key features of the out scattering are that the basic rate is determined by the particle-reservoir coupling and that the nonlinearity vanishes quadratically as $\omega_k\rightarrow 0$. The latter feature stems from the large charge gap and the vanishing of the charge-magnon coupling at low energies due to the Goldstone theorem.

This allows us to write down a kinetic equation that encodes magnon-magnon scattering through the collision integral $\mathcal{S}$ as well as the effects of drive and dissipation
\begin{align}
\partial_t n(\omega) =& \gamma_{\mathrm{in}}(1+n(\omega))-\gamma_{\mathrm{out}}\left(n(\omega)+\left(\frac{n(\omega)}{n_{\tilde{T}}(\omega)}\right)^2\right) \nonumber \\ &+\mathcal{S}\left[\{n(\omega)\}\right]\,.
\label{eq:interacting_differential_equation}
\end{align}

\subsection{Numerical implementation}
We discretize the system and solve the resulting set of coupled nonlinear equations numerically by integrating forward in time from an initial condition until a steady state is reached. We choose a uniform  $\ell\times \ell$ momentum space grid containing $N=\ell^2$ points shown in appendix~\ref{sec:pseudocode} and therefore a discrete set of momentum points $\omega_{\vect{k}}$. We replace all momentum/frequency integrals by sums. The largest linear dimension $\ell$ used throughout the paper is $\ell = 120$, which is the default discretization parameter for the results shown below, unless otherwise indicated. The discretized momentum grid is chosen in a way such that $\vect{k}=0$ is avoided because a Bose-Einstein distribution with $\mu = 0$ diverges as $\vect{k}\rightarrow 0$, implying that $\vect{k}=0$ cannot be treated directly numerically (see appendix~\ref{sec:pseudocode}, Figure~\ref{fig:Supplement_Grid}). Below we employ a careful finite-size scaling analysis and extrapolation to infinite system size to extract information about $\vect{k}\rightarrow 0$ and possible Bose-Einstein condensation.
In  the numerical results presented here we fix the parameter $\tilde{T}$ describing the nonlinear term in the dissipation to be  $\tilde{T}=0.6$  and set $\gamma_{\mathrm{out}}=0.002$, unless explicitly denoted otherwise. Our conclusions are independent of the specific parameter values. 

As noted above, the collision integral $\mathcal{S}$ conserves the magnon number $\mathcal{N}$ and energy $\mathcal{E}$ which are discretized as
\begin{subequations}\label{eq:Energy_Magnonnumber_Definition}
\begin{align}
\mathcal{N}&=\sum_{m=1}^{\omega_{\mathrm{max}}}
\rho\left(\omega_{m}\right)n\left(\omega_{m}\right)\\
\mathcal{E}&=\sum_{m=1}^{\omega_{\mathrm{max}}}
\rho\left(\omega_{m}\right)n\left(\omega_{m}\right)\omega_{m}\,,
\end{align}
\end{subequations}
where $\rho\left(\omega_{m}\right)$ is the discretization of the density of states given in Eq.~\eqref{eq:DOS_continous}
We parametrize the drive strength via the dimensionless tuning parameter, that controls the excitation density,
\begin{equation}
g\equiv\frac{\gamma_{\text{in}}}{\gamma_{\text{out}}}.\,,
\label{eq:g_definition}
\end{equation}
and consider the qualitative form of the computed magnon distribution function.

\section{Results}
\subsection{Nonequilibrium phase diagram}
Fig.~\ref{fig:1_Phase_diagram} summarizes our findings in terms of a phase diagram in the plane defined by the amplitude of quantum fluctuations (inverse spin length $1/S$, vertical axis) and the drive strength ($g$, horizontal axis). In equilibrium ($g=0$), increasing quantum fluctuations drives a transition to a quantum disordered state. Increasing the drive strength at a fixed value of quantum fluctuations produces two conceptually distinct effects. 

The drive adds energy to the system, exciting magnons above the ground state and thereby weakening the order. For drive strengths less than a critical value (here, $g=1$) the magnon distribution retains a subthermal form, with the magnon occupation $n(\omega)$ remaining finite as the magnon energy $\omega$ vanishes, in contrast to the $\sim T/\omega$ behavior of the thermal distribution. Although the distribution is subthermal, the increase in magnon number may be sufficient to drive the system into a disordered state, as indicated by the phase boundary in Fig.~\ref{fig:1_Phase_diagram}. This symmetry breaking phase transition is a nonequilibrium version of the standard equilibrium phase transition driven by raising temperature.  
Distinct from this transition Walldorf \textit{et al.}~also found a change in the magnon distribution from subthermal to superthermal, occurring as the relative drive strength was increased beyond the critical value $g=1$~\cite{walldorf_antiferromagnetic_2019}. It is this subthermal-to-superthermal transition, which is characterized by a qualitative change in the distribution and is not directly related to the disappearance of a conventional order parameter, that we investigate here. Because the distribution function is at least thermal, in the two-dimensional Heisenberg-symmetry case studied in detail here, long-range order is necessarily destroyed at $g=1$.
However, in two-dimensional xy/xxz or in three-dimensional systems, the ordered phase may persist into the superthermal phase. 

\begin{figure*}
	\centering
	\includegraphics[width= \textwidth]{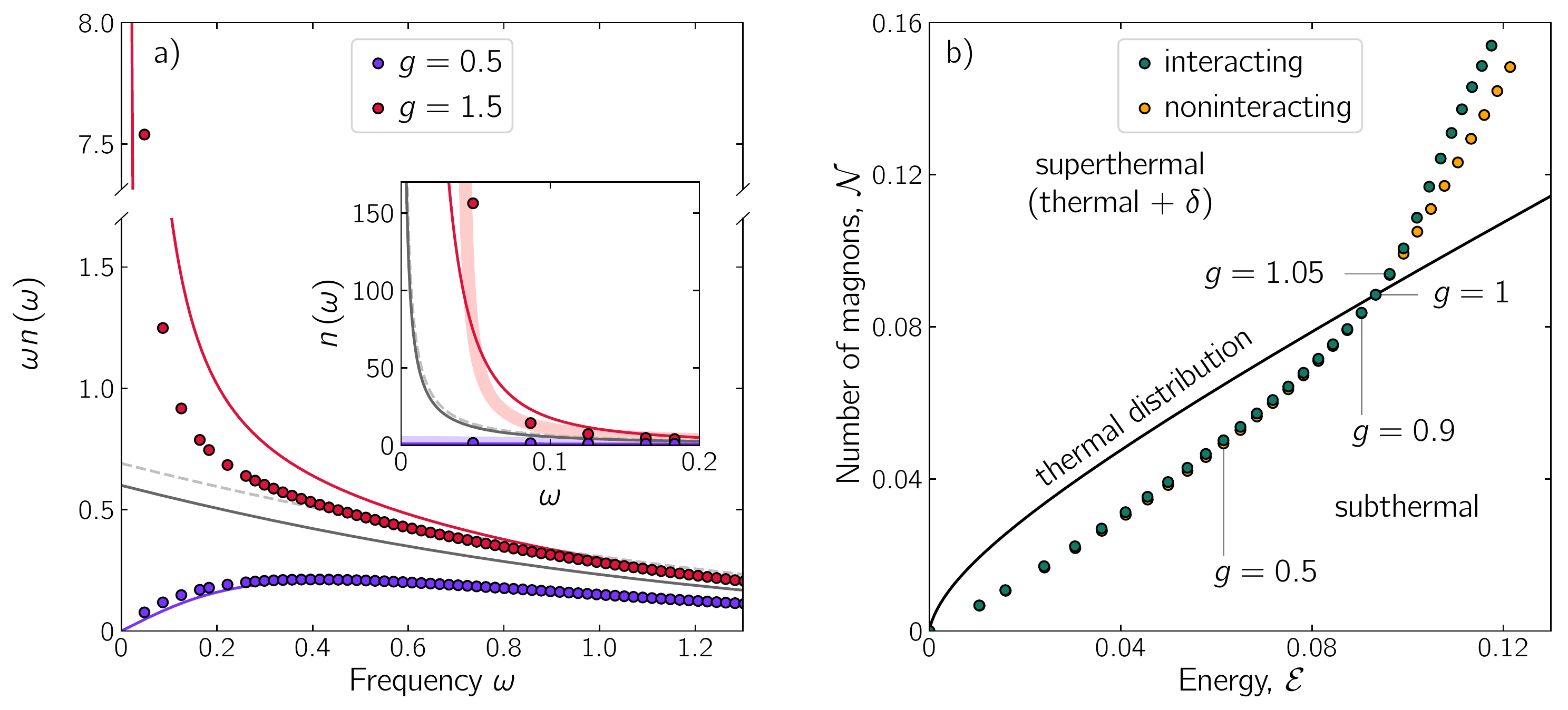}
	\caption{
	\textbf{Nonequilibrium phase transition.}
	a) Interaction-induced changes of steady-state magnon occupation $n(\omega)$. Plotted is $\omega n(\omega)$ as a function of magnon frequency $\omega$, in order to highlight the difference between subthermal ($\omega n(\omega) \rightarrow 0$ for $\omega \rightarrow 0$), thermal ($\omega n(\omega) \rightarrow \text{const}$), and superthermal ($\omega n(\omega) \rightarrow \infty$) regimes. The blue (red) data points show the interacting results for representative subthermal, $g=0.5$ (superthermal, $g=1.5$) cases, in comparison with the noninteracting results shown by blue (red) curves. The dark grey solid line indicates a thermal state at $g=1$ and $T=0.6$; the light grey dashed line is a best fit to the high-frequency part of the the interacting distribution function at $g=1.5$, and corresponds to a thermal state with an effective temperature $T > \tilde{T}$.
	Inset: The same results plotted as $n(\omega)$ versus $\omega$ focussing on the low-frequency part to highlight that the interacting superthermal system shows a low-frequency divergence that is stronger than both the noninteracting system and the best thermal fit.
	b)  {Points: magnon number vs total energy curve defined from  Eq.~\ref{eq:Energy_Magnonnumber_Definition} with $g$ as an implicit parameter for both noninteracting and interacting  steady state. Solid black line: magnon-number vs. 
	~total energy relation obtained from Bose distribution with chemical potential $\mu=0$ with temperature as implicit parameter. }  States below this critical Bose-Einstein condensation line have a lower number of magnons per energy than a thermal state. States above the critical line have a number of magnons that exceeds the maximal number in states with $\omega\left(\vect{k}\right)>0$ that is compatible with the given system energy in a thermal state, implying the existence of a $\delta$-function contribution at zero energy (condensate fraction) in the thermodynamic limit.
	}
	\label{fig:2_Steady_State_Occupations_Ratio_EN}
\end{figure*}

\subsection{Nonequilibrium steady state}
 Fig.~\ref{fig:2_Steady_State_Occupations_Ratio_EN} (a) compares the magnon distribution function calculated with and without magnon-magnon scattering. We find that the clear qualitative difference between the subthermal and superthermal cases is still evident in the interacting case, confirming that the nonequilibrium phase transition is preserved under magnon-magnon scattering. In the subthermal steady state, the impact of magnon-magnon scattering is rather small, producing only a slight shift of magnon occupation towards lower frequencies. In striking contrast, the superthermal steady state is strongly affected by magnon-magnon scattering. At all but the lowest frequency the effect of the scattering is to drive the distribution close to a thermal distribution, but the occupancy at the lowest frequency is strongly enhanced relative to the noninteracting case (see inset of Fig.~\ref{fig:2_Steady_State_Occupations_Ratio_EN} (a)). 

To interpret our results, we recall equilibrium BEC, where the occupancy is given by a Bose Einstein distribution with $\mu=0$ and a  $\delta$-function at $\omega_k=0$ describing the condensate fraction. This distribution has a temperature that is fixed by the total energy; the number of uncondensed bosons is then uniquely determined by this temperature, and any excess over the uncondensed number makes up the condensate fraction. With this in mind we plot in Fig.~\ref{fig:2_Steady_State_Occupations_Ratio_EN} (b) the magnon number as a function of magnon energy with $g$ as an implicit parameter, along with  the magnon number-energy relation implied by the Bose distribution with chemical potential $\mu=0$ and no condensate, with temperature as an implicit parameter. In ordinary BEC, decreasing the temperature decreases the energy moving the system to the left along a line at fixed $\mathcal{N}$. Crossing the solid line signals the BEC. In our system for $g<1$ the number-energy trace remains below the solid line. At $g=1$ the curves for both noninteracting and interacting systems cross the  solid line, implying for $g>1$ an excess of magnons. Importantly magnon-magnon interactions push the system even further away from the thermal distribution rather than towards it because magnon-magnon scattering tends to redistribute magnons towards lower energy, thus accommodating more magnons per energy compared to the noninteracting steady state.

\begin{figure}
	\centering
	\includegraphics[width= \columnwidth]{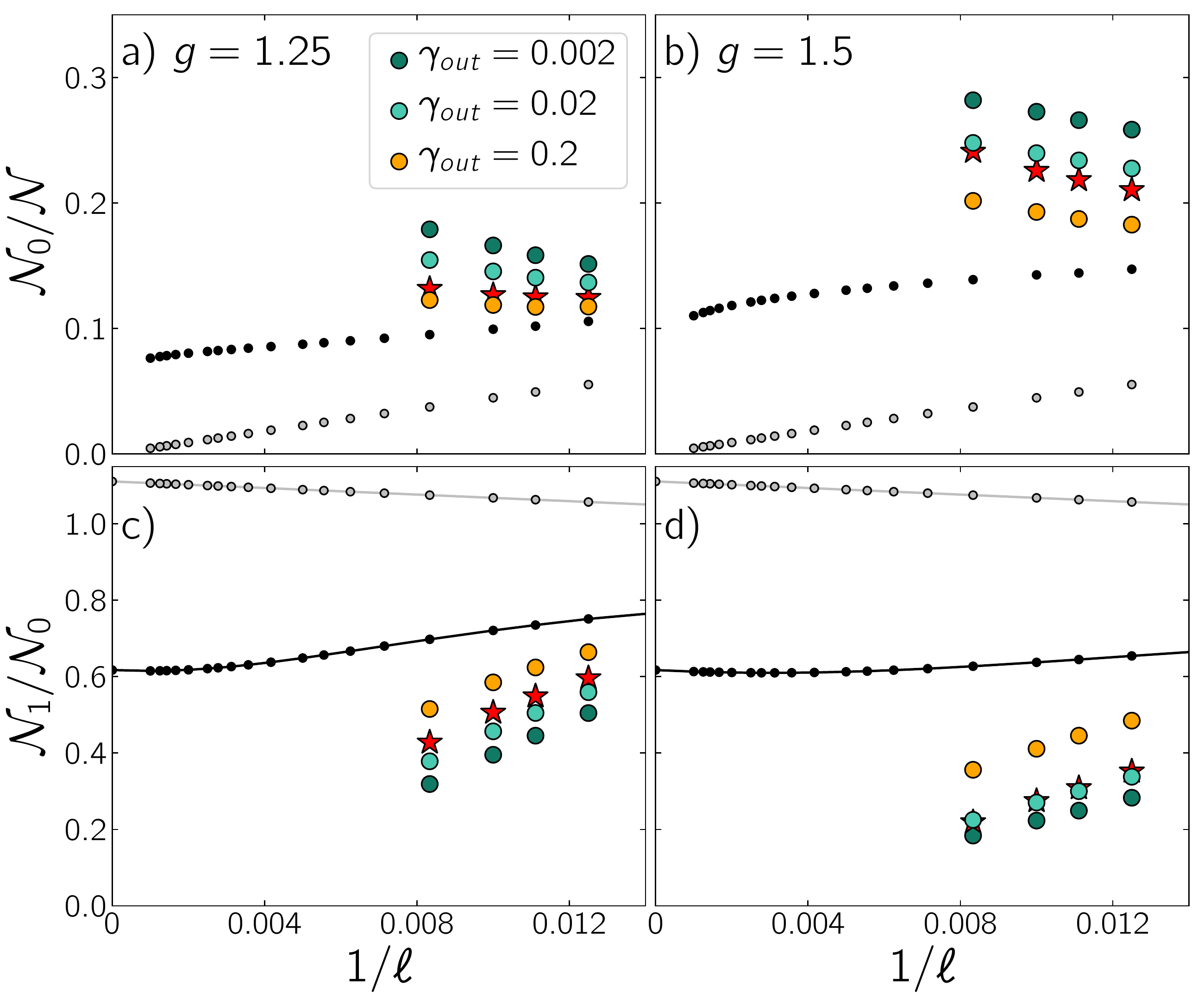}
	\caption{\textbf{Finite-size scaling analysis revealing $\delta$-function contribution at $g>1$ in the thermodynamic limit.} (a), (b) Ratio of the magnon density at the lowest frequency and the total number of magnons in the system, $\mathcal{N}_0/\mathcal{N}$, and (c), (d) ratio of the magnon density at the second lowest frequency and the lowest frequency, $\mathcal{N}_1/\mathcal{N}_0$, for $g=1.25$ (left panels) and $g=1.5$ (right panels). Different colors correspond to different values of $\gamma_{\text{out}}$ = $0.002, 0.02, 0.2$ as indicated. Black points correspond to the noninteracting stationary state, gray points show thermal behavior ($g=1$), and red stars correspond to the stationary state to which the interacting, closed system evolves when initialized with the respective noninteracting stationary state at given $g$. 
	}\label{fig:Magnon_Number_Scaling}
\end{figure}

\subsection{Finite size scaling analysis}
 To further interpret the data we present a finite-size scaling analysis. We define the magnon occupancy at the $m$-th frequency weighted by the discretized density of states, $\mathcal{N}_m = \rho(\omega_m) n(\omega_m)$.
 Fig.~\ref{fig:Magnon_Number_Scaling} (a) and (b) strongly suggest that the occupancy $\mathcal{N}_0$ of the the lowest frequency magnon mode remains a nonvanishing fraction of the overall number of magnons $\mathcal{N}$ as the system size increases in any interacting system with $g>1$. This is different from the case $g=1$, which has no condensate, and where the contribution of the lowest frequency vanishes as the system size increases. Fig.~\ref{fig:Magnon_Number_Scaling} (c) and (d) shows that the ratio of the occupancy at the second smallest frequency to the occupancy at the smallest frequency, $\mathcal{N}_1/\mathcal{N}_0$, is decreasing as the system size increases.  The decrease is apparently linear in $1/\ell$, but the system sizes available are not sufficient to allow for a precise determination.  The combination of a nonvanishing $\mathcal{N}_0/\mathcal{N}$ and a vanishing $\mathcal{N}_1/\mathcal{N}_0$ in the thermodynamic limit strongly suggests the existence of a $\delta$-function contribution at $\omega=0$. 
For reference, we also show data points for a system that is initialized with the noninteracting steady state at a given value of $g$ and then evolved as a closed system under magnon-magnon scattering. At $g>1$ this closed system is positioned above the critical line for BEC in the $\mathcal{N}$-$\mathcal{E}$ diagram in Fig.~\ref{fig:2_Steady_State_Occupations_Ratio_EN} (b). Therefore, in the thermodynamic limit this closed system necessarily develops a finite condensate fraction because this is the only possible thermalized solution to the closed-system kinetic equation. The comparison between the interacting driven-dissipative steady states and the closed-system thermalized states drives home our point that the interacting $g>1$ system develops a nonvanishing condensate fraction in the thermodynamic limit \footnote{For a corresponding analysis in the limit of weak driving, $g<1$, see SM.}.

\begin{figure}
	\centering
	\includegraphics[width= \columnwidth]{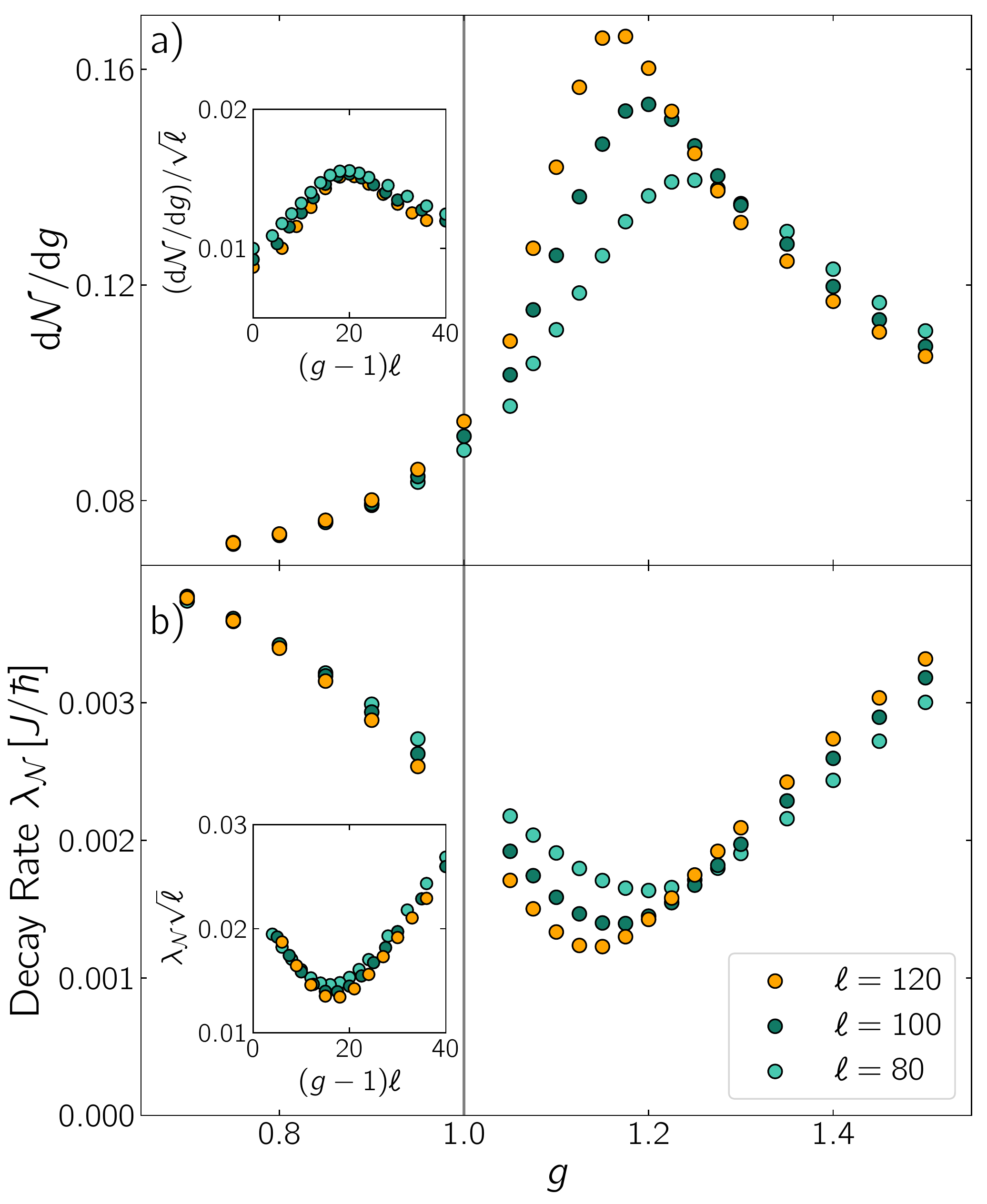}
	\caption{
	\textbf{Static and dynamical critical behavior in the interacting driven-dissipative steady state.} 
	(a) Rate of change of magnon number $\mathcal{N}$ as a function of $g$. Inset: Scaling behavior with linear system size collapses the data points onto a single curve. (b) Rate of decay of total magnon number $\mathcal{N}$ towards the stationary state, plotted as a function of $g$ for different system sizes as indicated. Inset: Scaling behavior with linear system size consistent with collapse onto a single curve, suggesting critical slowing down as $g\rightarrow 1$. 
	(for critical behavior in the strength of the condensate fraction, see SM)
	}
	\label{fig:Relaxation_constant_Energy_Particle_Number}
\end{figure}

\subsection{Static and dynamic criticality }
Fig.~\ref{fig:Relaxation_constant_Energy_Particle_Number} examines the nature of static and dynamic criticality occurring as $g$ is tuned through $g=1$.
The main panels show both the dependence of the static observable $\mathrm{d}\mathcal{N}/\mathrm{d}g$ [Fig.~\ref{fig:Relaxation_constant_Energy_Particle_Number} (a)] and the dynamic decay rate $\lambda_{\mathcal{N}}$ [Fig.~\ref{fig:Relaxation_constant_Energy_Particle_Number} (b)], as defined by
 \begin{align}
 \label{eq:DecayRate}
       \mathcal{N}\left(t\right)=\mathcal{N}_{\mathrm{final}}+\Delta\mathcal{N}\exp\left(-\lambda_{\mathcal{N}}t\right)
 \end{align}
 on the tuning parameter $g$. Equation~\eqref{eq:DecayRate} is the empirically observed long-time behavior of the excitation density in the system \footnote{We have checked that this slow time scale that emerges in the critical system is observed not only for the dynamics of the magnon number, but also for the total energy as well as the magnon occupation at any given energy.}. Data are shown for  different system sizes. For both quantities there is a clear difference between $g<1$ and $g>1$ with weak system-size dependence for $g<1$ and strong system-size dependence for $g>1$. The inset shows an approximate data collapse that is consistent with a critical scaling as $g\rightarrow 1$ from above and $\ell \rightarrow \infty$. The implication of the data collapse is that 
\begin{subequations}
\begin{align}
\label{eq:CriticalScaling}
      \frac{\mathrm{d}\mathcal{N}}{\mathrm{d}g} &= f_1\left[\left(g-1\right)\ell\right]\,\sqrt{\ell\,},
      \\
      \lambda_{\mathcal{N}} &= \frac{f_2\left[\left(g-1\right)\ell\right]}{\sqrt{\ell\,}}.
 \end{align} 
\end{subequations}
If $d\mathcal{N}/dg$ and $\lambda_N$ are to be finite and non-zero as $\ell \rightarrow \infty$, the two functions $f_1(x)$ and $f_2(x)$ need to have the form $f_1(x)\propto \left(1/\sqrt{x}\right)$ and $f_2(x)\propto \sqrt{x}$ as $x\rightarrow \infty$, implying that at $\ell =\infty$ $\mathrm{d}\mathcal{N}/\mathrm{d}g \sim \frac{1}{\sqrt{g-1}}$, i.e., a square-root singularity of $\mathcal{N}(g)$ in the thermodynamic limit,  and $\lambda_N\sim \sqrt{g-1}$ as $g \rightarrow 1^+$, i.e., a critical slowing down as $g\rightarrow 1$ from above. This asymmetric criticality is not present in the noninteracting theory and is a consequence of magnon-magnon interactions.


\section{Discussion}
A driven-dissipative system may exhibit two phase transitions as a function of drive strength. One is the nonequilibrium analogue of a conventional symmetry breaking transition, occurring because the drive creates excitations which push the system away from the ordered state. This transition has been previously studied \cite{Millis_2006_Nonequilibrium_Quantum_Criticality,Millis_2008_quantum_criticality_ferromagnets,sieberer_dynamical_2013, Sieberer_2014_Bose_condensation_polaritons,brennecke_real-time_2013,Rota_2019,BKTnoneq, Marino_Driven_Markovian_Quantum_Criticality_2016}. The other type, studied here, is that when the drive exceeds a critical value set by the linear dissipation mechanism, a kind of ``order from disorder'' transition may occur, with some fraction of the drive-induced excitations condensing into a zero-momentum ground state. Our finding bears an interesting relationship to the existing literature on Bose-Einstein condensation of magnons, where an evolution into a condensed state of a transiently induced magnon population is analysed. 

Crucial to our analysis is a numerically exact solution of the Boltzmann equation derived by considering the interactions among excitations, which enables an analysis of the interplay between the frequency dependence of the dissipation mechanism and the tendency to condensation. This comprehensive numerical solution extends previously published theory which typically uses either a phenomenological relaxation rate or a simple approximation to the magnon-magnon scattering term. A key finding is that the condensation occurs in a high-drive limit, where the drive induced energy density is large and the number of excited magnons is also large, and is associated with a dynamical (drive-strength driven) criticality. On the level of theory used here, this criticality is described by a new set of static and dynamic critical exponents. 

Our work raises many important questions. First, while we have demonstrated a qualitative change in the magnon distribution consistent with the formation of a condensate, the physics of fluctuations around this state has not yet been studied, and therefore a full analysis of the criticality, beyond the Boltzmann approximation used here, cannot be undertaken. Understanding how to characterize the differences between the nonthermal symmetry breaking transition and the usual thermal one, how to understand transitions involving distribution functions and not conventional order parameters, and how to generalize the standard equilibrium theory of spatial and temporal fluctuations in a critical state to strongly nonequilibrium situations such as that considered here, are important open problems. The issues are of particular importance in two dimensions, where the obvious generalization of the  Hohenberg-Mermin-Wagner theorem to nonequilibrium situations would suggest that the Bose-Einstein condensation we find signals a phase with power law correlations.

Observation of the nonthermal critical behavior predicted here is an important experimental challenge. Possible techniques include time-resolved second harmonic optical polarimetry or inelastic x-ray scattering \cite{mazzone_laser-induced_2021}. Our work also has a close connection to Bose-Einstein condensation in exciton-polariton systems, where interesting field-theory-based studies of criticality have appeared. \cite{sieberer_dynamical_2013,Sieberer_2014_Bose_condensation_polaritons}. Investigations of possible nonequilibrium-induced spatial structure, analogous to the structures observed in turbulence \cite{falkovich_lessons_2006}, and clarifying the relation of our work to nonthermal fixed points in closed systems after quenches \cite{berges_nonequilibrium_2015,erne_universal_2018, Demler_Universal_Prethermal_2020} are also important directions for future research.

\section{Acknowledgements}
We acknowledge discussions with S.~Diehl and M.~Mitrano.
This work was supported by the Max Planck-New York City Center for Nonequilibrium Quantum Phenomena.
MAS acknowledges financial support through the Deutsche Forschungsgemeinschaft (DFG, German Research Foundation) via the Emmy Noether program (SE 2558/2). DMK acknowledges support by the Deutsche
Forschungsgemeinschaft (DFG, German Research Foundation) via RTG 1995 and Germany’s Excellence Strategy -- Cluster of Excellence Matter and Light for Quantum Computing (ML4Q) EXC 2004/1 -- 390534769.   A.J.M. is supported in part by Programmable Quantum Materials, an Energy Frontier Research Center funded by the U.S. Department of Energy (DOE), Office of Science, Basic Energy Sciences (BES), under award DE-SC0019443. The Flatiron Institute is a division of the Simons Foundation.

\bibliographystyle{naturemag}
\bibliography{boltzmann,000_magnon_boltzmann}

\appendix

\section{Methods}\label{sec:methods}
\paragraph*{Interacting spin-wave theory}
We consider the isotropic Heisenberg antiferromagnet as given in Eq.~\eqref{eq:Heisenberg_Hamiltonian}
and apply standard Holstein-Primakoff spin-wave theory \cite{Holstein_Field_1940} resulting in 
\begin{align}
	H_{\mathrm{HP}}&= E_0 + H_0 + V\,.
\end{align}
with an irrelevant ground state energy $E_0$ and bilinear Hamiltonian
\begin{align}
	H_{0}
	&=
	\sum_{\kk}\hbar\,\omega_{\vect{k}}\left(\alpha_{\kk}^\dagger\alpha_{\kk}+ \beta_{\kk}^\dagger \beta_{\kk}\right).
\end{align}
The magnon dispersion is 
\begin{equation}\label{eq:Omega_isotropic}
    \omega_{\vect{k}}= \frac{J S z}{\hbar} 
    \left[1+\frac{1}{2S}\left(1-\frac{2}{N}\sum_{\kk'}\lambda_{\kk'}\right)\right]\lambda_{\kk}\,
\end{equation}
with
\begin{subequations}
\begin{align}
\lambda_{\kk}&=\sqrt{1-\gamma_{\kk}^2\,}\label{eq:lambda}\\
    \gamma_{\kk}
    &= \frac{\cos\left(k_x\right)+\cos\left(k_y\right)}{2}\,.
\end{align}
\end{subequations}
The interaction term for the kinematically allowed magnon energy and momentum conserving scattering processes is given by $V$ with interaction vertices $V_{1_{\alpha}^+ 2_{\alpha}^+ 3_{\alpha}^- 4_{\alpha}^-}^{(2:2)}$ and $\tilde{V}_{1_{\alpha}^+ 2_{\beta}^- 3_{\alpha}^- 4_{\beta}^+}^{(2:2)}$, namely
\begin{widetext}
\begin{subequations}
\begin{align}
\label{eq:scattering_processes}
	V &=
	-J\frac{2z}{N}
	\sum_{\kk_{1}\kk_{2}\kk_{3}\kk_{4}}
	\delta\left(\kk_{1}+\kk_{2}-\kk_{3}-\kk_{4}\right)
	\left\lbrace
	V^{(2:2)}\left(
	\alpha_1^\dagger \alpha_2^\dagger \alpha_3   \alpha_4
	+
	\beta_3^\dagger \beta_4^\dagger  \beta_1 \beta_2
	\right)
	+
	\tilde{V}^{(2:2)}
	\left(\alpha_1^\dagger \alpha_3  \beta_4^\dagger \beta_2\right)
	\right\rbrace
	\\
\label{eq:Vertices}
	V_{1_{\alpha}^+ 2_{\alpha}^+ 3_{\alpha}^- 4_{\alpha}^-}^{(2:2)}&=
	\gamma_{\left(2-4\right)}
	u_1 u_3 \, v_2 v_4 
	+\frac{1}{4}\left[
	\gamma_{1}\,
	u_1\, v_2  v_3 v_4 +
	\gamma_{2}\,
	u_1 u_3 u_4 \, v_2 +
	\gamma_{3}\,
	u_3\, v_1 v_2 v_4+
	\gamma_{4}\,
	u_1 u_2 u_3 \,v_4 \right]
	\\
	\tilde{V}_{1_{\alpha}^+ 2_{\beta}^- 3_{\alpha}^- 4_{\beta}^+}^{(2:2)}&=\gamma_{\left(2-4\right)}
	\left[
	u_1 u_2 u_3  u_4 + v_1 v_2 v_3 v_4 
	\right]
	+
	\gamma_{\left(2-3\right)}
	\left[
	u_1 u_2\, v_3 v_4 + u_3 u_4 \, v_1 v_2 
	\right]
	\\\nonumber
	&+
	\frac{1}{2}
	\gamma_{1}
	\left[
	u_3 \, v_1 v_2 v_4 + u_1 u_2 u_4\, v_3
	\right]
	+\frac{1}{2}
	\gamma_{2}
	\left[
	u_1 u_2 u_3 \,  v_4 + u_4\, v_1v_2  v_3
	\right]
	\\\nonumber
	&+\frac{1}{2}
	\gamma_{3}
	\left[
	u_2 u_3 u_4\, v_1 
	+
	u_1\, v_2 v_3 v_4 \right]
	+\frac{1}{2}
	\gamma_{4}
	\left[
	u_2\, v_1 v_3 v_4 + u_1 u_3 u_4  \,v_2
	\right]\,.
\end{align}\end{subequations}
\end{widetext}

Here we have used 
\begin{subequations}
\begin{align}
u_{\kk}&=\sqrt{\frac{1+\lambda_{\kk}}{2\lambda_{\kk}}}\\
v_{\kk}&=-\text{sign}\left(\gamma_{\kk}\right)\sqrt{\frac{1-\lambda_{\kk}}{2\lambda_{\kk}}}\,.
\end{align}
\end{subequations}
In Eq.~\eqref{eq:scattering_processes}, the momentum-conserving $\delta$ function is to be understood as modulo a reciprocal lattice vector of the standard two-dimensional antiferromagnetic Brillouin zone.

\paragraph*{Boltzmann equation} 
The semiclassical magnon Boltzmann equation for the magnon distribution in branch $\alpha$ at a given momentum $\kk_1$ is
\begin{align}
\frac{\mathrm{d}n^{\alpha}\left({\kk_1}\right)}{\mathrm{d}t}
=&
\frac{2\pi}{\hbar}
\left(\frac{2Jz}{N}\right)^2
\left(
	{\mathcal{S}}_\alpha^{\left(2:2\right)}\left(\kk_1\right)
	+ 
	\tilde{\mathcal{S}}_\alpha^{\left(2:2\right)}\left(\kk_1\right)
\right)
\end{align}
where $\mathcal{S}$ are the relevant scattering integrals. 
To leading order in $1/S$, only scattering processes with two magnons scattering into two other magnons are kinematically allowed. Consequently, the scattering conserves the number of magnons term by term at this level of approximation. 
The corresponding scattering integrals are given by
\begin{widetext}
\begin{align}\label{eq:SelfEnergy_aaaa}
	\mathcal{S}_\alpha^{\left(2:2\right)}\left(\kk_1\right)=&
	\sum_{\kk_2\kk_3\kk_4}
	\delta\left(\kk_1+\kk_2-\kk_3-\kk_4\right)
	\delta\left(\omega_{\kk_1}+\omega_{\kk_2}-\omega_{\kk_3}-\omega_{\kk_4}\right)
	\mathcal{V}^{\left(2:2\right)}_{1_{\alpha}^+ 2_{\alpha}^+ 3_{\alpha}^- 4_{\alpha}^-}
	\mathcal{V}^{\left(2:2\right)}_{3_{\alpha}^+ 4_{\alpha}^+ 1_{\alpha}^- 2_{\alpha}^-}
	\times 
	\\\nonumber
	&\qquad\qquad
	\left[
	\left(1+n^{\alpha}\left(\kk_1\right)\right)
	\left(1+n^{\alpha}\left(\kk_2\right)\right)
	n^{\alpha}\left(\kk_3\right)
	n^{\alpha}\left(\kk_4\right)
	-
	n^{\alpha}\left(\kk_1\right)
	n^{\alpha}\left(\kk_2\right)
	\left(1+n^{\alpha}\left(\kk_3\right)\right)
	\left(1+n^{\alpha}\left(\kk_4\right)\right)
	\right]
\end{align}
\begin{align}\label{eq:SelfEnergy_abab}
	\tilde{\mathcal{S}}_\alpha^{\left(2:2\right)}\left(\kk_1\right)=&
	\sum_{\kk_2\kk_3\kk_4}
	\delta\left(\kk_1+\kk_2-\kk_3-\kk_4\right)
	\delta\left(\omega_{\kk_1}+\omega_{\kk_2}-\omega_{\kk_3}-\omega_{\kk_3}\right)
	\tilde{\mathcal{V}}^{\left(2:2\right)}_{1_{\alpha}^+ 2_{\beta}^- 3_{\alpha}^- 4_{\beta}^+}
	\tilde{\mathcal{V}}^{\left(2:2\right)}_{3_{\alpha}^+ 4_{\beta}^- 1_{\alpha}^- 2_{\beta}^+}
	\times 
	\\\nonumber
	&\qquad\qquad
	\left[
	\left(1+n^{\alpha}\left(\kk_1\right)\right)
	\left(1+n^{\beta}\left(\kk_4\right)\right)
	n^{\alpha}\left(\kk_3\right)
	n^{\beta}\left(\kk_2\right)
	-
	n^{\alpha}\left(\kk_3\right)
	n^{\beta}\left(\kk_2\right)
	\left(1+n^{\alpha}\left(\kk_1\right)\right)
	\left(1+n^{\beta}\left(\kk_4\right)\right)
	\right]\,.
\end{align}
\end{widetext} 

\paragraph*{Computational remarks} 
We compute the time evolution on the two-dimensional antiferromagnetic Brillouin zone, that is discretized into square tiles and subsequently mapped onto an energy grid (see appendix~\ref{sec:pseudocode} for details). The time propagation of the full kinetic equation in the main text is performed using the two-step Adams–Bashforth method. We have carefully checked convergence in the time step discretization.

The staggered magnetization is computed via  
\begin{align}\label{eq:magnetization}
        m\left(S, n(\omega)\right)
		&=
		S + \frac{1}{2} -\omega_{\mathrm{max}}
		\sum_{m=1}^{\omega_{\mathrm{max}}}
		\frac{\rho\left(\omega_{m}\right)}{\omega_{m}}\left(n\left(\omega_{m}\right)+\frac{1}{2}\right).
\end{align}
Specifically, the black curve in Fig.~\ref{fig:1_Phase_diagram} that separates the subthermal disordered phase from the subthermal ordered phase is computed by solving the equation $m\left(S, n(\omega)\right)=0$ (with the noninteracting magnon distribution at given $g$ inserted to compute $m$) for $1/S$.

\section{Strength of the condensate fraction}
\begin{figure}[h!]
	\centering
	\includegraphics[width= 0.85\columnwidth]{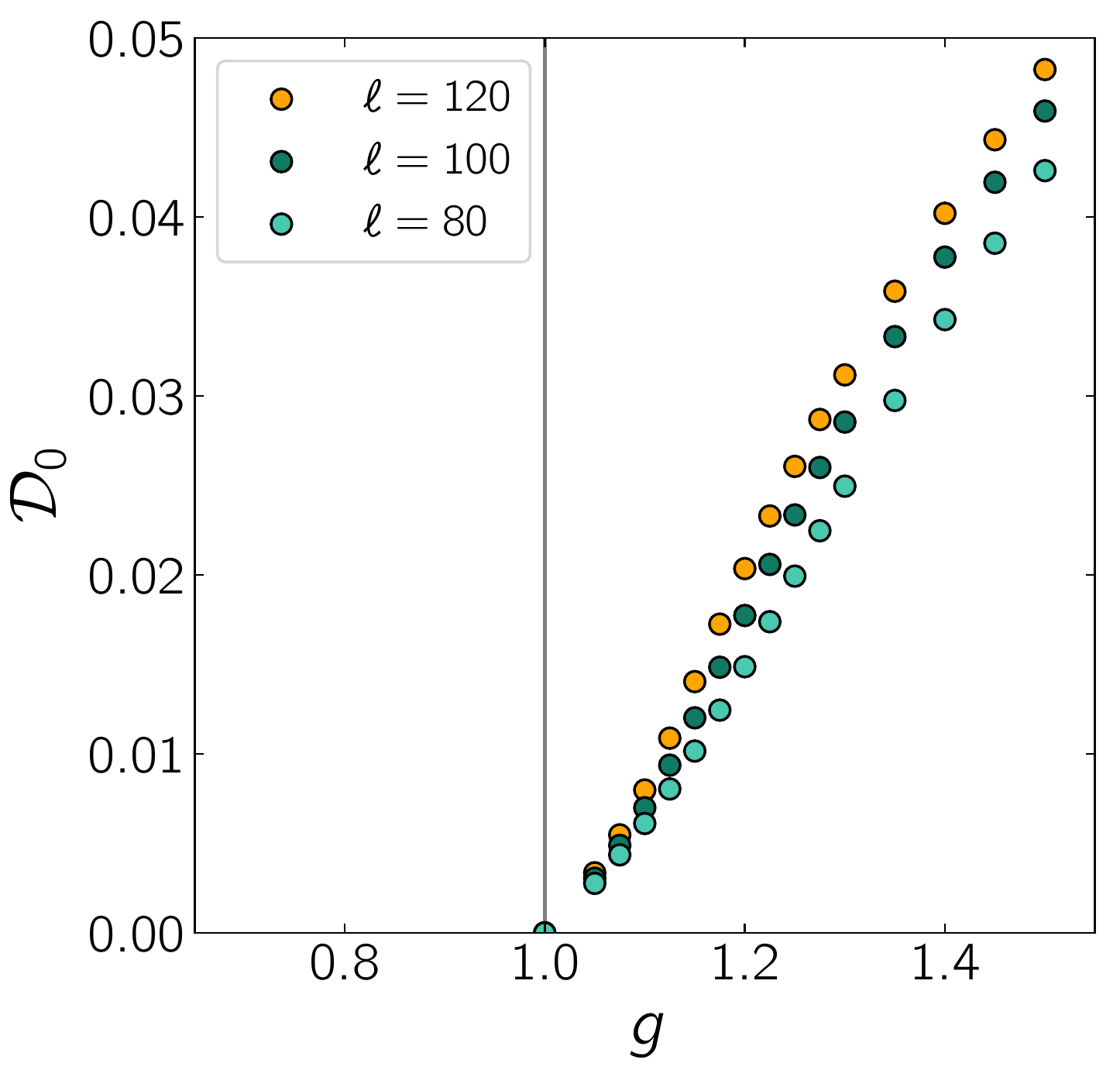}
	\caption{Condensate fraction $\mathcal{D}_0$ as a function of dimensionless tuning parameter $g$ for different linear system sizes as indicated. 
	}\label{fig:Supplement_Delta_Strength}
\end{figure}

The strength of the condensate faction is determined by the ratio of the number of magnons $\mathcal{N}$ to the system energy $\mathcal{E}$. Projecting each individual point in Fig. 2b) vertically onto the thermal distribution gives the number of magnons $\mathcal{N}_{\mathrm{th}}$ that can be accommodated by the thermal distribution. The excess of magnons determines the strength of the delta function, $\mathcal{D}_0 \equiv \mathcal{N}-\mathcal{N}_{\mathrm{th}}$. Therefore the steady state has the form 
\begin{equation}
   \mathcal{N}\left(\omega\right)=\mathcal{D}_0\,\mathcal{N}_{\Delta}\left(\omega\right) +\mathcal{N}_{\mathrm{th},T_{\mathcal{E}}} \left(\omega\right), 
\end{equation}
where $\mathcal{N}_{\Delta}\left(\omega\right)$ is a normalized function (integrating to unity) and, as discussed above, turns into a $\delta$-function in the thermodynamic limit. Since the number of magnons only exceeds the number of magnons in a thermal distribution at $g>1$, the weight of the $\delta$-function $\mathcal{D}_0$ vanishes for $g<1$. 
The decrease of the weight of the delta function $\mathcal{D}_0$ to $0$ at $g=1$ is marking the phase transition [Fig.~\ref{fig:Supplement_Delta_Strength}].

\section{Scaling of the magnon number in the limit of weak driving}

In the low driving phase $g<1$ the scaling behavior is substantially different from the results in the strong drive phase. As it is visible in Fig.~\ref{fig:Supplement_Scaling} a) where the  contribution  of  the  lowest  frequency in the interacting phase  goes  to  zero  as  system  size  is  increased, just as in the thermal system. So at $g<1$ there are no indications for a condensate fraction at $\omega=0$. Similarly, there is only a minimal shift from the non-interacting results in the ratio of the magnon density at the second lowest frequency and the lowest frequency, $\mathcal{N}_1/\mathcal{N}_0$. This is behavior in the low drive ordered phase is substantially different from the findings in the high drive, disordered phase.

\begin{widetext}

\begin{figure*}[h!]
	\centering
	\includegraphics[width= 0.85\textwidth]{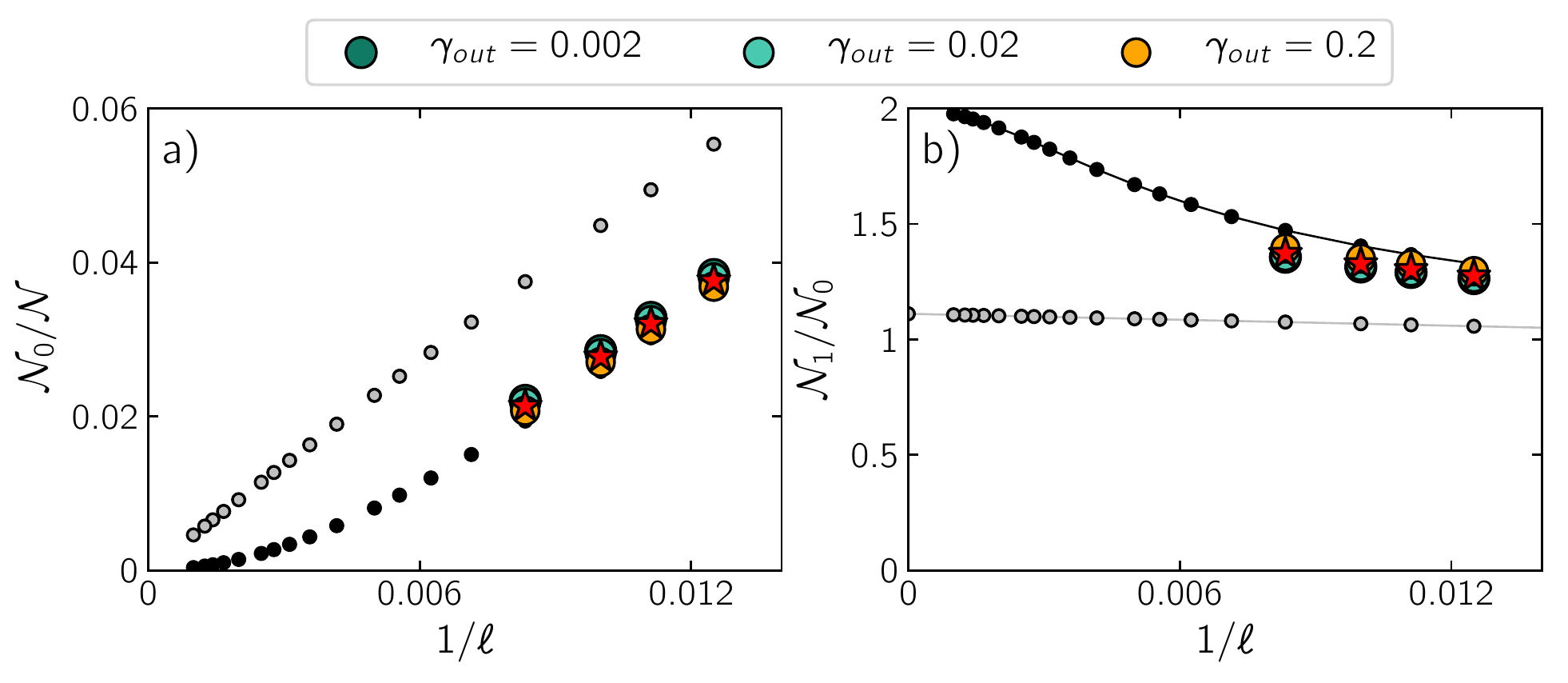}
	\caption{Finite-size scaling analysis analogous to Fig. 4 for the subthermal regime ($g=0.875$).
	  (a) Ratio of the magnon density at the lowest frequency and the total number of magnons in the system, $\mathcal{N}_0/\mathcal{N}$, and (b) ratio of the magnon density at the second lowest frequency and the lowest frequency, $\mathcal{N}_1/\mathcal{N}_0$. Different colors correspond to different values of $\gamma_{\text{out}}$ = $0.002, 0.02, 0.2$ as indicated. Black points correspond to the non-interacting stationary state, gray points show thermal behavior ($g=1$), and red stars correspond to the stationary state to which the interacting, closed system evolves when initialized with the respective non-interacting stationary state at given $g$. 
	}\label{fig:Supplement_Scaling}
\end{figure*}

\section{Pseudocode}
\label{sec:pseudocode}

We numerically consider a quadratic lattice of momentum vectors as displayed in Fig~\ref{fig:Supplement_Grid} with linear dimension $\ell$ and $\ell^2$ lattice sites. To make our compuation numerically feasible even for comparatively large $\ell$ we then reduce this MBZ using symmetry relations to $\left(\ell^2+2\ell\right)/8$ lattice sites (green). These reduced MBZ vectors ($k_{\mathrm{PZ}}$) are associated with different weights due to their multiplicity as indicated.  
Please note that in the following pseudocode $\#$ denotes the number of a quantity in an array while names like $\vect{k}_{\mathrm{PZ}}$ without a $\#$ are the actual quantity. For example $\vect{k}_{\mathrm{PZ}}$ without a $\#$ is the actual vector in the reduced MBZ.
\begin{algorithmic}[1]
\Statex\hrulefill
\Statex 1) Building the full (yellow) and reduced (green) MBZ as displayed in Fig. \ref{fig:Supplement_Grid} 
\Statex\hrulefill
    \State Save MBZ vectors sorted by length in $\mathrm{MBZ}\left[\mathrm{\#}\vect{k}_{\mathrm{MBZ}}\right]\left[k_x,k_y\right]$\;
    \State Save vectors within the reduced MBZ sorted by length in  $\mathrm{PZ}\left[\mathrm{\#}\vect{k}_{\mathrm{PZ}}\right]\left[k_x,k_y\right]$\;
		\For {$k\in\mathrm{PZ}$}
			\State save the precise Energy associated with this vector as $\Omega\left[\mathrm{\#}\vect{k}_{\mathrm{PZ}}\right]$
			\State save the weight associated with this vector as $\mathrm{kweight}\left[\mathrm{\#}\vect{k}_{\mathrm{PZ}}\right]$
		\EndFor
		\Statex\hrulefill
\end{algorithmic}

The scattering conserves both momentum and energy. This is implemented numerically by mapping the MBZ in momentum space on an energy grid as displayed in Fig.~\ref{fig:Supplement_Energy_Grid}. To do so, we divide the interval $\left\lbrace 0, \Omega_{\mathrm{max}}\right\rbrace$ into $\ell$ equidistant energy bins and determine with which bin the vectors in the momentum grid are associated. The different colors of the bins in Fig.~\ref{fig:Supplement_Energy_Grid} are simply to distinguish them from each other and have no further meaning. Since not all bins will have energies not all bins need to be taken into account. Note that in the example of $\ell=8$ only 5 of the bis are occupied (purple $\omega$). Each bin is then associated with the total weight of the MBZ vectors in it (red numbers).
\begin{figure}[h!]
	\centering
	\includegraphics[width= 0.6\columnwidth]{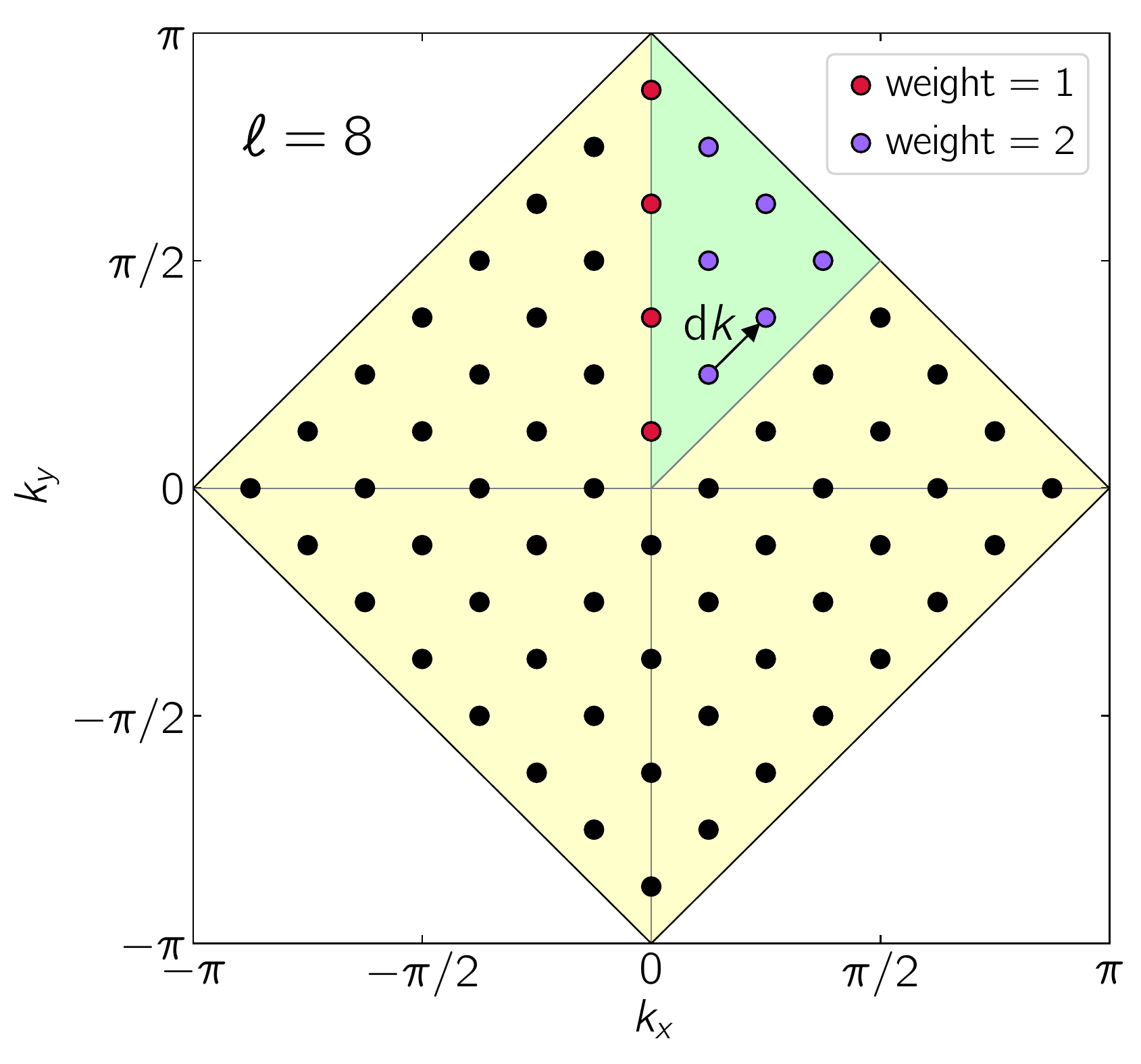}
	\caption{Magnetic Brillouin zone (MBZ) for $\ell=8$. The full MBZ (yellow) can be reduced to $\left(\ell^2+2\ell\right)/8$ lattice sites (green) due to the symmetry of the lattice. The multiplicity weights of these reduced lattice vectors that are sufficient to simulate the dynamics in the system is marked as indicated. 
	}\label{fig:Supplement_Grid}
\end{figure}

\begin{algorithmic}[1]
\Statex\hrulefill
\Statex 2) Map the reduced MBZ in $\vect{k}$ space onto an energy grid as illustrated in Fig.~\ref{fig:Supplement_Energy_Grid}
\Statex\hrulefill
    \State Divide the interval $\left\lbrace 0, \Omega_{\mathrm{max}}\right\rbrace$ into $\ell$ equidistant energy bins 
    \Statex(see Fig.~\ref{fig:Supplement_Energy_Grid}, blue and magenta boxes)
    \For {$k\in\mathrm{PZ}$}
        \State identify in which energy bin $\Omega\left[\# k\right]$ falls 
    \EndFor
    \State Discard empty energy bins 
    \State Save the center of the remaining energy bins as $\mathrm{energybin}\left[\#\omega\right]$ 
    \Statex(see purple $\left\lbrace\omega_1,\omega_2,\omega_3,\omega_4,\omega_5\right\rbrace$ in Fig.~\ref{fig:Supplement_Energy_Grid})
    \State Save the numbers of the reduced MBZ vectors in each bin as $\mathrm{kpz}@\mathrm{energybin}\left[\#\omega\right]\left[\#k_{\mathrm{PZ}}\right]$
    \State Compute the total kweight in each bin and save it as $\mathrm{kweight}@\mathrm{energybin}\left[\#\omega\right]\left[\#k_{\mathrm{PZ}}\right]$
    \Statex\hrulefill
\end{algorithmic}

The next step is to find the quadruples in momentum space that satisfy momentum and energy conservation simultaneously. Note that we use the centers of the energy bins and not the precise energies of the momentum vectors to determine weather energy conservation is satisfied.  The factor $4$ in the cutoff is needed because each quadruple consists of $4$ momentum vectors. Furthermore, all entries of the $2$ dimensional array "integrals" are the same. Here the cutoff has to be divided by $\ell^4$ because there are 4 free dimensions in the integration. The vertices are then symmetrized by computing 
\begin{align}
\left(\mathcal{VV}\right)_{\mathrm{sym}}
    &=
    0.125
    \left[
    \mathcal{V}_{1_{\alpha}^+ 2_{\alpha}^+ 3_{\alpha}^-4_{\alpha}^-}
    +\mathcal{V}_{3_{\alpha}^+ 4_{\alpha}^+ 1_{\alpha}^-2_{\alpha}^-}
    \right]
    +0.125
    \left[
    \mathcal{V}_{1_{\alpha}^+ 2_{\alpha}^+ 4_{\alpha}^-3_{\alpha}^-}
    +\mathcal{V}_{3_{\alpha}^+ 4_{\alpha}^+ 2_{\alpha}^-1_{\alpha}^-}
    \right]
    \\\nonumber
    &\quad
    +0.125
    \left[
    \mathcal{V}_{2_{\alpha}^+ 1_{\alpha}^+ 3_{\alpha}^-4_{\alpha}^-}
    +\mathcal{V}_{4_{\alpha}^+ 3_{\alpha}^+ 1_{\alpha}^-2_{\alpha}^-}
    \right]
    +0.125
    \left[
    \mathcal{V}_{2_{\alpha}^+ 1_{\alpha}^+ 4_{\alpha}^-3_{\alpha}^-}
    +\mathcal{V}_{4_{\alpha}^+ 3_{\alpha}^+ 2_{\alpha}^-1_{\alpha}^-}
    \right]
\end{align}
and 
\begin{align}
    \left(\tilde{\mathcal{V}}\tilde{\mathcal{V}}\right)_{\mathrm{sym}}
    &=
    0.25
    \left[
    \tilde{V}_{1_{\alpha}^+ 4_{\beta}^- 3_{\alpha}^- 2_{\beta}^+}
    +
    \tilde{V}_{3_{\alpha}^+ 2_{\beta}^- 1_{\alpha}^- 4_{\beta}^+}
    \right]
    +
    0.25
    \left[
    \tilde{V}_{2_{\alpha}^+ 3_{\beta}^- 4_{\alpha}^- 1_{\beta}^+}
    +
    \tilde{V}_{4_{\alpha}^+ 1_{\beta}^- 2_{\alpha}^- 3_{\beta}^+}
    \right]
\end{align}
This vertex symmetrization ensures energy- and particle number conservation by enforcing detailed balance and is a necessary step in the energy-grid-representation. 

\begin{algorithmic}[1]
\Statex\hrulefill
\Statex 3) Find Quadruples that satisfy momentum and energy conservation in momentum space
\Statex\hrulefill
\State $\mathrm{cutoff}=4*\Omega_{\mathrm{max}}/\ell$
    \For{$\vect{k}_1\in \mathrm{PZ}$}
        \For{$\vect{k}_2\in \mathrm{MBZ}$}
            \For{$\vect{k}_3\in \mathrm{MBZ}$}
                \State $\vect{k}_4=\vect{k}_1+\vect{k}_2-\vect{k}_{3}$
                \State Find bin energy $\omega_{i}$ associated with each of $\left\lbrace\vect{k}_1,\vect{k}_2,\vect{k}_3,\vect{k}_4\right\rbrace
                \rightarrow\left\lbrace\omega_1,\omega_2,\omega_3,\omega_4\right\rbrace$
                \If{$\omega_1+\omega_2-\omega_3-\omega_4<0.05*\mathrm{cutoff}$}
                    \State Save quadruple as  $\mathrm{kquadruple}\left[\#k1\right]\left[\#\mathrm{quadruple}\right]\left[\left\lbrace\vect{k}_1,\vect{k}_2,\vect{k}_3,\vect{k}_4\right\rbrace\right]$
                    \State Compute $\left(\mathcal{VV}\right)_{\mathrm{sym}}=\left(\mathrm{symmetrize}
                    \left[\mathcal{V}_{1_{\alpha}^+ 2_{\alpha}^+ 3_{\alpha}^-4_{\alpha}^-}
                    \right]\right)^2$
                    \State Compute $\left(\tilde{\mathcal{V}}\tilde{\mathcal{V}}\right)_{\mathrm{sym}}=\left(\mathrm{symmetrize}\left[
                    \tilde{V}_{1_{\alpha}^+ 2_{\beta}^- 3_{\alpha}^- 4_{\beta}^+}
                    \right]\right)^2$
                    \State Set $\mathrm{vertices}\left[\#k_1\right] \left[\#\mathrm{quadruple}\right] = 
                    \mathcal{VV}_{\mathrm{sym}}+\tilde{\mathcal{V}}\tilde{\mathcal{V}}_{\mathrm{sym}}$
                    \State Set $\mathrm{integrals} \left[\#k_1\right] \left[\#\mathrm{quadruple}\right] = \mathrm{cutoff}/(\ell^4) $
                \EndIf
            \EndFor
        \EndFor
    \EndFor
    \Statex\hrulefill
\end{algorithmic}

\begin{figure*}[b]
	\centering
	\includegraphics[width= 0.8\columnwidth]{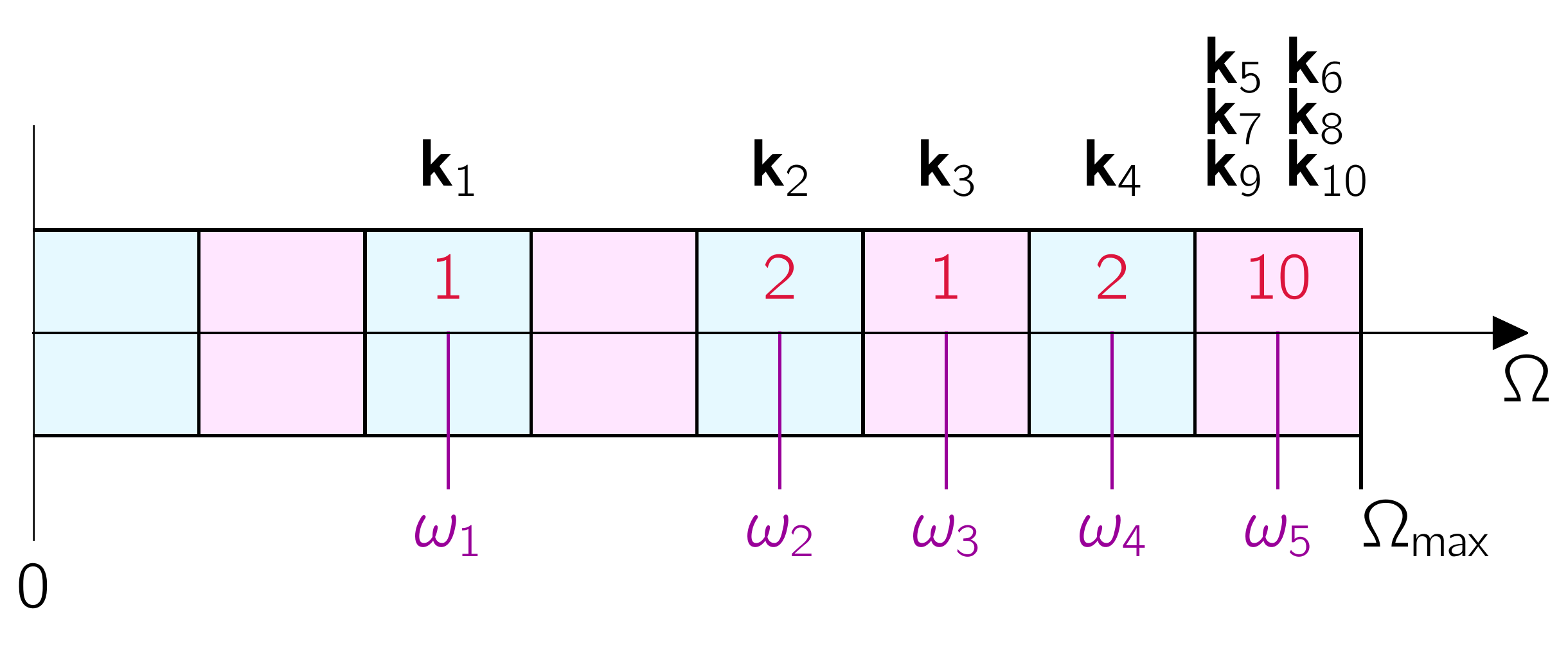}
	\caption{Mapping of momentum grid onto an energy grid for $\ell=8$. The Interval $\left\lbrace 0, \Omega_{\mathrm{max}}\right\rbrace$ into $\ell$ equidistant energy bins (blue and magenta) and for each momentum vector $\vect{k}_\mathrm{PZ}$ the associated bin is determined. The red numbers give the total weight of all vectors within the energy bin, so for example the energy bin $\omega_5$ has the momentum vectors $\left\lbrace\vect{k}_5,\vect{k}_6,\vect{k}_7,\vect{k}_8,\vect{k}_9,\vect{k}_{10}\right\rbrace$ that have a total weight of 10.}
	\label{fig:Supplement_Energy_Grid}
\end{figure*}
Now we have found the quadruples in momentum space, but in order to compute the time evolution using the energy grid in Fig.~\ref{fig:Supplement_Energy_Grid} we need to turn the quadruple list into an energy list with $\omega_1$, $\omega_2$, $\omega_3$ and $\omega_4$ and then average for each given $\omega_1$ over the multiple entries. This gives a consolidated list of energy quadruples and their weights. 

\begin{algorithmic}[1]
\Statex\hrulefill
\Statex 4) Convert Momentum Quadruples into Energy Quadruples
\Statex\hrulefill
    \For{$\omega\in\mathrm{energybins}$}
        \For{$k\in\mathrm{kweight}@\mathrm{energybin}\left[\#\omega\right]$}
            \For{$q\in\mathrm{kquadruple}\left[\#k1\right]$}
                \State $\left\lbrace k_1, k_2, k_3, k_4\right\rbrace= \mathrm{kquadruple} [\# k] [\# q]$
                \State Find energy bins associated with $k_1$, $k_2$, $k_3$ and $k_4$ $\rightarrow$ 
                $\left\lbrace \omega_1, \omega_2, \omega_3, \omega_4\right\rbrace$
                \State Safe $\mathrm{energyquadruples}\left[\#\omega\right]\left[\#\mathrm{equadruple}\right]
                \left[\left\lbrace  \omega_1, \omega_2, \omega_3, \omega_4 \right\rbrace\right]$
                \State $\mathrm{enegryweight}=
                \mathrm{integrals}\left[\# k\right]\left[\# q\right]*\mathrm{vertices}\left[\# k\right]\left[\# q\right]*\mathrm{kweight}\left[\# k\right]$
                \State Set $\mathrm{energyweights}\left[\#\omega\right]\left[\#\mathrm{equadruple}\right]=
                \mathrm{energyweight}$
            \EndFor
        \EndFor
        \For{$\mathrm{equad}\in \mathrm{energyquadruples}\left[\#\omega\right]$}
            \State Check if the combination $\left\lbrace \omega_1, \omega_2, \omega_3, \omega_4\right\rbrace$ has already been found
            \If{No}
                \State Save $\mathrm{energyquadruples}\_\mathrm{consolidated}\left[\#\omega\right]
                \left[\#\mathrm{equad}\_\mathrm{c}\right] \left[\left\lbrace  \omega_1, \omega_2, \omega_3, \omega_4 \right\rbrace\right]$
                \State $\mathrm{energyweight}\_\mathrm{averaged}=\mathrm{energyweights}\left[\#\omega\right]/\mathrm{kweight}@\mathrm{energybin}\left[\#\omega\right]$
                \State Save $\mathrm{energyweights}\_\mathrm{consolidated}\left[\#\omega\right] \left[\#\mathrm{equad}\_\mathrm{c}\right]=\mathrm{energyweight}\_\mathrm{averaged}$
            \ElsIf{Yes}
                \State $\mathrm{energyweight}\_\mathrm{averaged}=\mathrm{energyweights}\left[\#\omega\right]/\mathrm{kweight}@\mathrm{energybin}\left[\#\omega\right]$
                \State Add $\mathrm{energyweights}\_\mathrm{consolidated}\left[\#\omega\right] \left[\#\mathrm{equad}\_\mathrm{c}\right]+=\mathrm{energyweight}\_\mathrm{averaged}$
            \EndIf
        \EndFor
    \EndFor
    \Statex\hrulefill
\end{algorithmic}
We then use the consolidated quadruples in energy space to compute the time evolution using the  two-step Adams–Bashforth linear multistep method.
\end{widetext}


\end{document}